\documentclass[10pt,a4paper,onecolumn]{article}
\usepackage{marginnote}
\usepackage{graphicx}
\usepackage{xcolor}
\usepackage{authblk,etoolbox}
\usepackage{titlesec}
\usepackage{calc}
\usepackage{tikz}
\usepackage{hyperref}
\hypersetup{colorlinks,breaklinks,
            urlcolor=[rgb]{0.0, 0.5, 1.0},
            linkcolor=[rgb]{0.0, 0.5, 1.0}}
\usepackage{caption}
\usepackage{tcolorbox}
\usepackage{amssymb,amsmath}
\usepackage{ifxetex,ifluatex}
\usepackage{seqsplit}
\usepackage{xstring}

\usepackage{fixltx2e} 
\usepackage[
  backend=biber,
]{biblatex}
\bibliography{paper.bib}

\let\textttOrig=\texttt
\def\texttt#1{\expandafter\textttOrig{\seqsplit{#1}}}
\renewcommand{\seqinsert}{\ifmmode
  \allowbreak
  \else\penalty6000\hspace{0pt plus 0.02em}\fi}

\makeatletter
\let\href@Orig=\href
\def\href@Urllike#1#2{\href@Orig{#1}{\begingroup
    \def\Url@String{#2}\Url@FormatString
    \endgroup}}
\def\href@Notdoi#1#2{\def\tempa{#1}\def\tempb{#2}%
  \ifx\tempa\tempb\relax\href@Urllike{#1}{#2}\else
  \href@Orig{#1}{#2}\fi}
\def\href#1#2{%
  \IfBeginWith{#1}{https://doi.org}%
  {\href@Urllike{#1}{#2}}{\href@Notdoi{#1}{#2}}}
\makeatother

\usepackage[top=3.5cm, bottom=3cm, right=1.5cm, left=1.0cm,
            headheight=2.2cm, reversemp, includemp, marginparwidth=4.5cm]{geometry}

\titleformat{\section}
  {\normalfont\sffamily\Large\bfseries}
  {}{0pt}{}
\titleformat{\subsection}
  {\normalfont\sffamily\large\bfseries}
  {}{0pt}{}
\titleformat{\subsubsection}
  {\normalfont\sffamily\bfseries}
  {}{0pt}{}
\titleformat*{\paragraph}
  {\sffamily\normalsize}

\usepackage{fancyhdr}
\makeatletter
\let\ps@plain\ps@fancy
\fancyheadoffset[L]{4.5cm}
\fancyfootoffset[L]{4.5cm}

\definecolor{linky}{rgb}{0.0, 0.5, 1.0}

\newtcolorbox{repobox}
   {colback=red, colframe=red!75!black,
     boxrule=0.5pt, arc=2pt, left=6pt, right=6pt, top=3pt, bottom=3pt}

\newcommand{\ExternalLink}{%
   \tikz[x=1.2ex, y=1.2ex, baseline=-0.05ex]{%
       \begin{scope}[x=1ex, y=1ex]
           \clip (-0.1,-0.1)
               --++ (-0, 1.2)
               --++ (0.6, 0)
               --++ (0, -0.6)
               --++ (0.6, 0)
               --++ (0, -1);
           \path[draw,
               line width = 0.5,
               rounded corners=0.5]
               (0,0) rectangle (1,1);
       \end{scope}
       \path[draw, line width = 0.5] (0.5, 0.5)
           -- (1, 1);
       \path[draw, line width = 0.5] (0.6, 1)
           -- (1, 1) -- (1, 0.6);
       }
   }

\patchcmd{\@maketitle}{center}{flushleft}{}{}
\patchcmd{\@maketitle}{center}{flushleft}{}{}
\patchcmd{\@maketitle}{\LARGE}{\LARGE\sffamily}{}{}
\def\maketitle{{%
  
  \AB@maketitle}}
\makeatletter
\renewcommand\AB@affilsepx{ \protect\Affilfont}
\renewcommand\AB@affilnote[1]{{\bfseries #1}\hspace{3pt}}
\renewcommand{\affil}[2][]%
   {\newaffiltrue\let\AB@blk@and\AB@pand
      \if\relax#1\relax\def\AB@note{\AB@thenote}\else\def\AB@note{#1}%
        \setcounter{Maxaffil}{0}\fi
        \begingroup
        \let\href=\href@Orig
        \let\texttt=\textttOrig
        \let\protect\@unexpandable@protect
        \def\thanks{\protect\thanks}\def\footnote{\protect\footnote}%
        \@temptokena=\expandafter{\AB@authors}%
        {\def\\{\protect\\\protect\Affilfont}\xdef\AB@temp{#2}}%
         \xdef\AB@authors{\the\@temptokena\AB@las\AB@au@str
         \protect\\[\affilsep]\protect\Affilfont\AB@temp}%
         \gdef\AB@las{}\gdef\AB@au@str{}%
        {\def\\{, \ignorespaces}\xdef\AB@temp{#2}}%
        \@temptokena=\expandafter{\AB@affillist}%
        \xdef\AB@affillist{\the\@temptokena \AB@affilsep
          \AB@affilnote{\AB@note}\protect\Affilfont\AB@temp}%
      \endgroup
       \let\AB@affilsep\AB@affilsepx
}
\makeatother

\renewcommand\Affilfont{\sffamily\small\mdseries}
\ifnum 0\ifxetex 1\fi\ifluatex 1\fi=0 
  \usepackage[T1]{fontenc}
  \usepackage[utf8]{inputenc}

\else 
  \ifxetex
    \usepackage{mathspec}
  \else
    \usepackage{fontspec}
  \fi
  \defaultfontfeatures{Ligatures=TeX,Scale=MatchLowercase}

\fi
\IfFileExists{upquote.sty}{\usepackage{upquote}}{}
\IfFileExists{microtype.sty}{%
\usepackage{microtype}
\UseMicrotypeSet[protrusion]{basicmath} 
}{}

\usepackage{hyperref}
\hypersetup{unicode=true,
            pdftitle={fgivenx: A Python package for functional posterior plotting},
            pdfborder={0 0 0},
            breaklinks=true}
\let\addcontentslineOrig=\addcontentsline
\def\addcontentsline#1#2#3{\bgroup
  \let\texttt=\textttOrig\addcontentslineOrig{#1}{#2}{#3}\egroup}
\let\markbothOrig\markboth
\def\markboth#1#2{\bgroup
  \let\texttt=\textttOrig\markbothOrig{#1}{#2}\egroup}
\let\markrightOrig\markright
\def\markright#1{\bgroup
  \let\texttt=\textttOrig\markrightOrig{#1}\egroup}

\usepackage{graphicx,grffile}
\makeatletter
\def\maxwidth{\ifdim\Gin@nat@width>\linewidth\linewidth\else\Gin@nat@width\fi}
\def\maxheight{\ifdim\Gin@nat@height>\textheight\textheight\else\Gin@nat@height\fi}
\makeatother
\setkeys{Gin}{width=\maxwidth,height=\maxheight,keepaspectratio}
\IfFileExists{parskip.sty}{%
\usepackage{parskip}
}{
\setlength{\parindent}{0pt}
\setlength{\parskip}{6pt plus 2pt minus 1pt}
}
\ifx\paragraph\undefined\else
\let\oldparagraph\paragraph
\renewcommand{\paragraph}[1]{\oldparagraph{#1}\mbox{}}
\fi
\ifx\subparagraph\undefined\else
\let\oldsubparagraph\subparagraph
\renewcommand{\subparagraph}[1]{\oldsubparagraph{#1}\mbox{}}
\fi

\title{fgivenx: A Python package for functional posterior plotting}

        \author[1, 2, 3]{Will Handley}
    
      \affil[1]{Astrophysics Group, Cavendish Laboratory, J.J.Thomson Avenue, Cambridge,
CB3 0HE, UK}
      \affil[2]{Kavli Institute for Cosmology, Madingley Road, Cambridge, CB3 0HA, UK}
      \affil[3]{Gonville \& Caius College, Trinity Street, Cambridge, CB2 1TA, UK}
  \date{\vspace{-5ex}}

\begin{document}
\maketitle

\marginpar{
  \sffamily\small

  {\bfseries DOI:} \href{https://doi.org/10.21105/joss.00849}{\color{linky}{10.21105/joss.00849}}

  \vspace{2mm}

  {\bfseries Software}
  \begin{itemize}
    \setlength\itemsep{0em}
    \item \href{https://github.com/openjournals/joss-reviews/issues/849}{\color{linky}{Review}} \ExternalLink
    \item \href{https://github.com/williamjameshandley/fgivenx}{\color{linky}{Repository}} \ExternalLink
    \item \href{10.5281/zenodo.1404584}{\color{linky}{Archive}} \ExternalLink
  \end{itemize}

  \vspace{2mm}

  {\bfseries Submitted:} 18 July 2018\\
  {\bfseries Published:} 28 August 2018

  \vspace{2mm}
  {\bfseries License}\\
  Authors of papers retain copyright and release the work under a Creative Commons Attribution 4.0 International License (\href{http://creativecommons.org/licenses/by/4.0/}{\color{linky}{CC-BY}}).
}

\hypertarget{summary}{%
\section{Summary}\label{summary}}

Researchers are often concerned with numerical values of parameters in
numerical models. Our knowledge of such things can be quantified and
presented using probability distributions as demonstrated in Figure 1.

\begin{figure}
\centering
\includegraphics{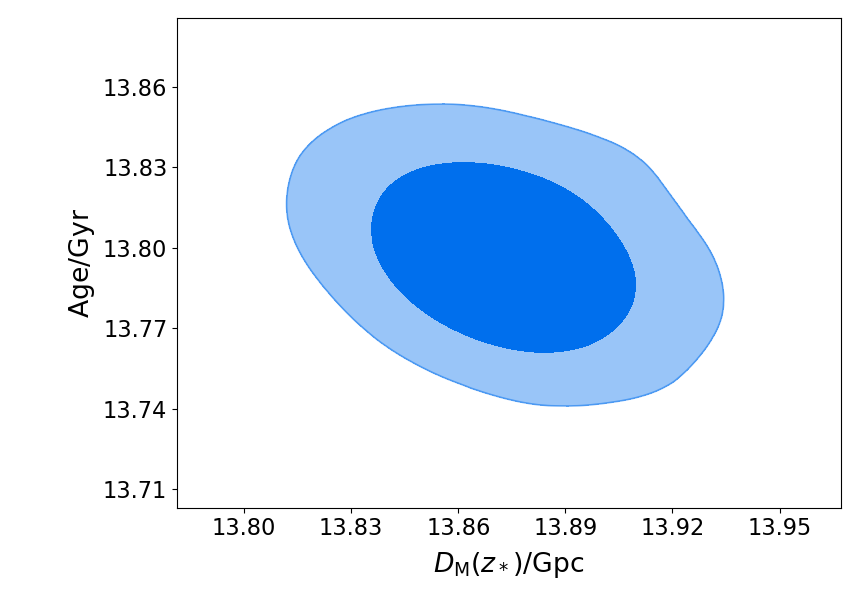}
\caption{The age and size of the universe, as measured using Planck 2018
data. (non-Astro)Physicists may note that 14 Gigaparsecs is roughly 46
billion light years. The fact that the observable universe is roughly
three times larger in light years in comparison with its age is
explained by the expansion of space over cosmic history. Contours
indicate 67\% and 95\% marginalised iso-probability credibility
regions.}
\end{figure}

Contour plots such as Figure 1 can be created using two-dimensional
kernel density estimation using packages such as
\href{https://docs.scipy.org/doc/scipy/reference/generated/scipy.stats.gaussian_kde.html}{scipy}
(Jones, Oliphant, and Peterson 2001),
\href{http://getdist.readthedocs.io/en/latest/intro.html}{getdist}
(Lewis 2015), \href{https://corner.readthedocs.io/en/latest/}{corner}
(Foreman-Mackey 2016) and
\href{https://pygtc.readthedocs.io/en/latest/}{pygtc} (Bocquet and
Carter 2016), where the samples provided as inputs to such programs are
typically created by a Markov Chain Monte Carlo (MCMC) analysis. For
further information on MCMC and Bayesian analysis in general,
``Information Theory, Inference and Learning Algorithms'' is highly
recommended (MacKay 2002), which is available freely
\href{http://www.inference.org.uk/itprnn/book.html}{online}.

As well as quantifying the uncertainty of real-valued parameters,
scientists may also be interested in producing a probability
distribution for the predictive posterior of a function \texttt{f(x)}.
Take as a universally-relatable case the equation of a line
\texttt{y\ =\ m*x\ +\ c}. Given posterior probability distributions for
the gradient \texttt{m} and intercept \texttt{c}, then the ability to
predict \texttt{y} knowing \texttt{x} given their linear relationship
would also be characterized by some uncertainty. This is depicted as
\texttt{P(y\textbar{}x)} in the bottom right panel of Figure 2.

\begin{figure}
\centering
\includegraphics{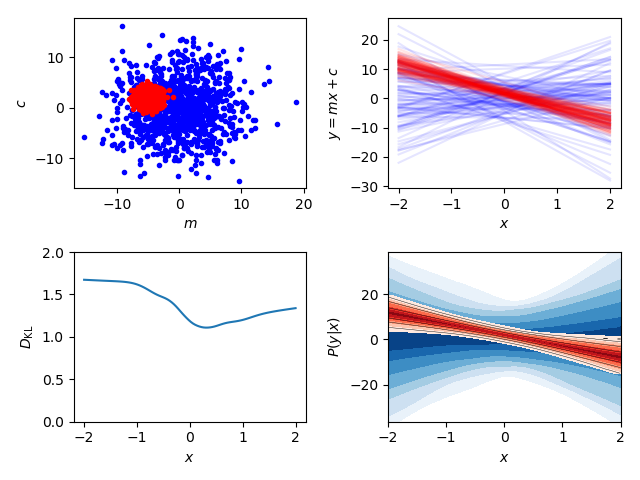}
\caption{Example output of fgivenx provided some prior and posterior
samples and a linear test function. Top-left: underlying parameter
covariances between \texttt{m} and \texttt{c} for realizations from the
prior (blue) and from the posterior (red). Top-right realisations
function \texttt{y=m*x+c}. Bottom-left: The conditional Kullback-Leibler
divergence. Bottom-right: The probability of measuring y for a given x,
essential a contour version of the panel directly above, where contours
indicate 67\%, 95\% and 99\% iso-probability credibility regions.}
\end{figure}

\texttt{fgivenx} is a Python package for showing the relationships as
depicted in Figure 2, including the conditional Kullback-Leibler
divergence (Kullback and Leibler 1951). This \texttt{y=m*x+c} example
provides a simple illustration, but the code has been used in recent
Planck studies to quantify our knowledge of the primordial power
spectrum of curvature perturbations (Planck Collaboration 2016)(Planck
Collaboration 2018a)(Planck Collaboration 2018b), in examining the dark
energy equation of state (Hee et al. 2016) (Hee et al. 2017) for
measuring errors in parameter estimation (Higson et al. 2017), for
providing diagnostic tests for nested sampling (Higson et al. 2018b) and
for Bayesian compressive sensing (Higson et al. 2018a).

\texttt{fgivenx} is a Python package for functional posterior plotting,
currently used in astronomy, but will be of use to scientists performing
any Bayesian analysis which has predictive posteriors that are
functions. The source code for \texttt{fgivenx} is available on
\href{https://github.com/williamjameshandley/fgivenx}{GitHub} and has
been archived as \texttt{v2.1.17} to Zenodo with the linked DOI:
(Handley 2018).

\hypertarget{acknowledgements}{%
\section{Acknowledgements}\label{acknowledgements}}

Contributions and bug-testing were provided by Ed Higson and Sonke Hee.

\hypertarget{references}{%
\section*{References}\label{references}}
\addcontentsline{toc}{section}{References}

\hypertarget{refs}{}
\leavevmode\hypertarget{ref-pygtc}{}%
Bocquet, Sebastian, and Faustin W. Carter. 2016. ``Pygtc: Beautiful
Parameter Covariance Plots (Aka. Giant Triangle Confusograms).''
\emph{The Journal of Open Source Software} 1 (6).
\url{https://doi.org/10.21105/joss.00046}.

\leavevmode\hypertarget{ref-corner}{}%
Foreman-Mackey, Daniel. 2016. ``Corner.py: Scatterplot Matrices in
Python.'' \emph{The Journal of Open Source Software} 24.
\url{https://doi.org/10.21105/joss.00024}.

\leavevmode\hypertarget{ref-zenodo}{}%
Handley, W. 2018. ``Fgivenx: V2.1.17,'' August.
\url{https://doi.org/10.5281/zenodo.1404584}.

\leavevmode\hypertarget{ref-Hee2015}{}%
Hee, S., W. J. Handley, M. P. Hobson, and A. N. Lasenby. 2016.
``Bayesian model selection without evidences: application to the dark
energy equation-of-state.'' \emph{MNRAS} 455 (January): 2461--73.
\url{https://doi.org/10.1093/mnras/stv2217}.

\leavevmode\hypertarget{ref-Hee2016}{}%
Hee, S., J. A. Vázquez, W. J. Handley, M. P. Hobson, and A. N. Lasenby.
2017. ``Constraining the dark energy equation of state using Bayes
theorem and the Kullback-Leibler divergence.'' \emph{MNRAS} 466 (April):
369--77. \url{https://doi.org/10.1093/mnras/stw3102}.

\leavevmode\hypertarget{ref-Higson2018b}{}%
Higson, E., W. Handley, M. Hobson, and A. Lasenby. 2018a. ``Bayesian
Sparse Reconstruction: A Brute-Force Approach to Astronomical Imaging
and Machine Learning.''

\leavevmode\hypertarget{ref-Higson2017}{}%
---------. 2017. ``Sampling Errors in Nested Sampling Parameter
Estimation.'' \emph{ArXiv E-Prints}, March.
\url{http://arxiv.org/abs/1703.09701}.

\leavevmode\hypertarget{ref-Higson2018}{}%
---------. 2018b. ``Diagnostic Tests for Nested Sampling Calculations.''
\emph{ArXiv E-Prints}, April. \url{http://arxiv.org/abs/1804.06406}.

\leavevmode\hypertarget{ref-scipy}{}%
Jones, Eric, Travis Oliphant, and Pearu Peterson. 2001. ``SciPy: Open
Source Scientific Tools for Python.'' \url{http://www.scipy.org/}.

\leavevmode\hypertarget{ref-Kullback}{}%
Kullback, S., and R. A. Leibler. 1951. ``On Information and
Sufficiency.'' \emph{Ann. Math. Statist.} 22 (1): 79--86.
\url{https://doi.org/10.1214/aoms/1177729694}.

\leavevmode\hypertarget{ref-getdist}{}%
Lewis, Anthony. 2015. ``Getdist Github Repository.''
\url{https://github.com/cmbant/getdist}.

\leavevmode\hypertarget{ref-mackay}{}%
MacKay, David J. C. 2002. \emph{Information Theory, Inference \&
Learning Algorithms}. New York, NY, USA: Cambridge University Press.

\leavevmode\hypertarget{ref-inflation2015}{}%
Planck Collaboration. 2016. ``Planck 2015 results. XX. Constraints on
inflation.'' \emph{A\&A} 594 (September): A20.
\url{https://doi.org/10.1051/0004-6361/201525898}.

\leavevmode\hypertarget{ref-legacy2018}{}%
---------. 2018a. ``Planck 2018 results. I. Overview and the
cosmological legacy of Planck.'' \emph{ArXiv E-Prints}, July.
\url{http://arxiv.org/abs/1807.06205}.

\leavevmode\hypertarget{ref-inflation2018}{}%
---------. 2018b. ``Planck 2018 results. X. Constraints on inflation.''
\emph{ArXiv E-Prints}, July. \url{http://arxiv.org/abs/1807.06211}.

\end{document}